\documentclass[twocolumn,aps,prb,superscriptaddress,10pt]{revtex4-2}
\usepackage{amsmath,amssymb,mathrsfs}
\usepackage{natbib}
\usepackage{subfigure}
\usepackage{tabularx}
\usepackage{epsfig}
\usepackage{longtable}
\usepackage{amsfonts}
\usepackage{rotating}
\usepackage{bbold}
\usepackage{hhline}
\usepackage{braket}
\usepackage{txfonts, comment}

\usepackage[unicode=trueb,bookmarks=true,bookmarksnumbered=false,bookmarksopen=false,breaklinks=false,pdfborder={0 0 1},backref=false,colorlinks=true]{hyperref}

\hypersetup{linkcolor=magenta,urlcolor=blue,citecolor=blue,pdfstartview={FitH},hyperfootnotes=false,unicode=true}

\def\be{\begin{equation}}
\def\ee{\end{equation}}
\def\bea{\begin{eqnarray}}
\def\eea{\end{eqnarray}}

\begin{document}
\title{Complexity enriched dynamical phases for fermions on graphs}

\author{Wei Xia}
\affiliation{Shanghai Qi Zhi Institute, AI Tower, Xuhui District, Shanghai 200232, China}
\affiliation{Shanghai Artificial Intelligence Laboratory, Shanghai 200232, China} 

\author{Jie Zou}
\affiliation{State Key Laboratory of Surface Physics, Institute of Nanoelectronics and Quantum Computing,
and Department of Physics, Fudan University, Shanghai 200433, China
}
\affiliation{Hefei National Laboratory, Hefei 230088, China}  
\affiliation{Shanghai Branch, Hefei National Laboratory, Shanghai 201315, China}  
\author{Xiaopeng Li}
\email{xiaopeng\_li@fudan.edu.cn}
\affiliation{State Key Laboratory of Surface Physics, Institute of Nanoelectronics and Quantum Computing,
and Department of Physics, Fudan University, Shanghai 200433, China
}
\affiliation{Shanghai Qi Zhi Institute, AI Tower, Xuhui District, Shanghai 200232, China}
\affiliation{Shanghai Artificial Intelligence Laboratory, Shanghai 200232, China} 
\affiliation{Shanghai Research Center for Quantum Sciences, Shanghai 201315, China}

\begin{abstract}
Dynamical quantum phase transitions, encompassing phenomena like many-body localization transitions and measurement-induced phase transitions, are often characterized and identified through the analysis of quantum entanglement. Here, we highlight that the dynamical phases defined by entanglement are further enriched by complexity. We investigate the entanglement, Krylov dimension and Krylov complexity for fermions on regular graphs, which can be implemented by systems like $^6$Li atoms confined by optical tweezers. Our investigations unveil that while entanglement follows volume laws on both types of regular graphs with degree $d = 2$ and $d = 3$, the Krylov complexity exhibits distinctive behaviors, with Krylov complexity and Krylov dimension sharing the same scaling relationship. We analyze both free fermions and interacting fermions models. In the absence of interaction, both numerical results and theoretical analysis confirm that the dimension of the Krylov space scales as $D\sim N$ for regular graphs of degree $d = 2$ with $N$ sites, and we have $D\sim N^2$ for $d = 3$. The qualitative distinction also persists in interacting fermions on regular graphs. For interacting fermions, our theoretical analyses find the dimension scales as $D\sim 4^{N^\alpha}$ for regular graphs of $d = 2$ with  $0.38\leq\alpha\leq0.59$, whereas it scales as $D\sim 4^N$ for $d = 3$. The distinction in the complexity of quantum dynamics for fermions on graphs with different connectivity can be probed in experiments by measuring the out-of-time-order correlators. 
\end{abstract}

\date{\today}
\maketitle

\section{Introduction}
Dynamical quantum phase transitions stand apart from conventional phase transitions by their dependence on the passage of time rather than on control parameters like temperature or pressure~\cite{Heyl_2018,Zvyagin_2016,Heyl_2013,Heyl_2014,Vosk_2014,Heyl_2015,Jurcevic_2017}. These transitions manifest across a spectrum of systems, with many-body localization transitions serving as a prototypical example~\cite{Nandkishore_2015, Abanin_2019, Pal_2010,alet_2018}. In such transitions, the system progresses from an ergodic phase to a many-body localized phase with increasing disorder strength. Another notable example is measurement-induced phase transitions~\cite{Skinner_2019,koh2023measurement,Choi_2020, Gullans_2020}, where the system exhibits two dynamical phases: entangling and disentangling phases. Investigations into dynamical quantum phase transitions also extend to topological systems. For instance, phase transitions consistently emerge in quenches between two Hamiltonians with differing absolute values of the Chern number~\cite{Vajna2015, Huang2016, Budich2016, Bhattacharya2017, flaschner2018}. 
Recent advancements in this field encompass the identification of dynamical order parameters~\cite{Budich2016,Sharma2016,flaschner2018,Bhattacharya2017Mixed, Bhattacharya2017}, exploration of scaling and universality~\cite{Heyl2015Scaling,Karrasch2017Dynamical}, and culminate in experimental observations~\cite{flaschner2018,Jurcevic2017Direct,Choi2016Exploring}.

Over the past two decades, quantum platforms have made significant strides in controlling physical systems at the quantum level. Dynamical phase transitions have been observed in various systems including ultracold atoms~\cite{muniz2020exploring,Scott2019Observation}, trapped ions~\cite{Tian2020Observation}, superconductors~\cite{google2023measurement,koh2023measurement,dborin2022simulating,Kai2020Probing}, and Rydberg atoms~\cite{Bluvstein2021Controlling,Dong2024Ergodicity,wu2023observation,liu2024bifurcation}. Among these, programmable Rydberg atom arrays have garnered attention due to their scalability, flexible interactions, and versatility~\cite{Bluvstein2021Controlling,Omran2019Generation,scholl2021quantum,bluvstein2022quantum,bornet2023scalable,browaeys2020many}. Recent studies have utilized programmable Rydberg atom arrays to explore graph problems such as maximum independent set~\cite{Ebadi2022Quantum,kim2022rydberg}, maximum cut~\cite{goswami2023solving,Zhou2020Quantum}, and maximum weighted independent set~\cite{Nguyen2023Quantum,Wild2021Quantum}. One remarkable progress accomplished in the last few years is atoms can be freely transported across remote sites with optical tweezer techniques~\cite{barredo2016atom,endres2016atom}, which permits investigations of fascinating quantum many-body dynamics on graphs beyond conventional paradigms that largely focus on locally interacting models, or non-local models with a specific type of long-range interactions.

\begin{figure}[htp]
\includegraphics[width=.85\linewidth]{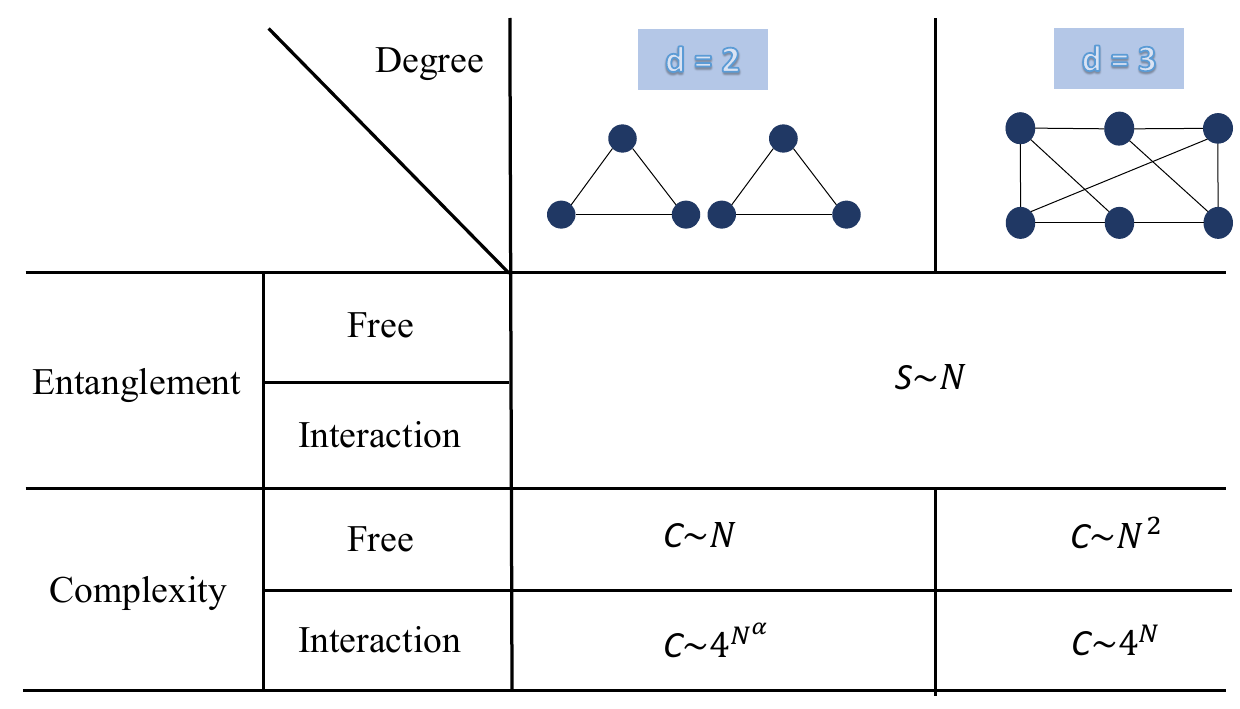}
\caption{The scaling laws of entanglement and complexity for regular graph models with degrees $d = 2$ and $d = 3$. In all scenarios, entanglement adheres to volume laws, i.e., $S\sim N$. We unveil rich dynamical phases in these models by examining the Krylov complexity. 
For non-interacting fermions, the Krylov dimension $C$ with $d=2$ scales as $C\sim N$, whereas for $d = 3$, it scales as $C\sim N^2$. 
 In the interacting case, the scalings of the Krylov dimension for $d=2$, and $d=3$ are $C\sim 4^{N^{\alpha}}$ ($0.38 \leq \alpha \leq 0.59$), and  $C\sim 4^N$, respectively.
 }
\label{fig:SketchMap}
\end{figure}

Entanglement typically serves as a distinguishing factor between different phases in dynamical phase transitions. In the context of many-body localization transitions, ergodic phases are characterized by linear growth of entanglement, whereas localization phases exhibit logarithmic growth~\cite{Abanin_2019}. Similar phenomena are observed in measurement-induced phase transitions, where linear growth denotes entangling phases and logarithmic growth denotes disentangling phases~\cite{Skinner_2019}. In our study, we observe the emergence of new phases characterized by Krylov complexity, as illustrated in Fig~\ref{fig:SketchMap}. Krylov complexity is proposed as a novel indicator for evaluating operator growth more directly and quantitatively~\cite{Parker2019Universal,bhattacharjee2022krylov}. It has received tremendous research efforts in various contexts such as quantum chaos and integrable systems~\cite{rabinovici2021operator,rabinovici2022krylov,rabinovici2022krylov,hashimoto2023krylov,Fabian2022Krylov,Kim2022Operator}, quantum field theory~\cite{barbon2019evolution,Dymarsky2021Krylov,avdoshkin2022krylov,kundu2023state}, and open systems~\cite{bhattacharya2022operator,bhattacharjee2023operator,bhattacharya2023krylov,bhattacharjee2024operator,Liu2023Krylov}.
We investigate Krylov complexity for a half-filled free fermion model on regular graphs with $ N $ sites, and the results are summarized in Fig ~\ref{fig:SketchMap}. Notably, the quantum entanglement on regular graphs with degrees $ d=2 $ and $ d=3 $ both follows the volume law, with entanglement entropy $ S $ scaling linearly as $ \sim N $. This indicates that these models belong to the same dynamical phase from the perspective of entanglement. However, their Krylov complexity exhibits different scaling behaviors: $ C $ scales as $ \sim N $ for degree $ d = 2 $ and $ \sim N^2 $ for $d = 3$. To clarify the quantitative differences between regular graphs of degrees $ d = 2 $ and $ d = 3 $, we develop a theory that accurately captures the scaling laws using the Krylov dimension. We also extend our analysis to include interaction effects. In the interacting case, the Krylov dimension grows exponentially with system size, making it challenging for numerical methods to precisely determine the scaling law exponent. By extending our theory from the free case to the interacting case, we find that the scaling laws shift to $ C \sim 4^{N^{\alpha}} $. Specifically, we find $ 0.38 \leq \alpha \leq 0.59 $ for $ d = 2 $, whereas $ \alpha = 1$ for $ d = 3 $. Thus, these models belong to different dynamical phases from the viewpoint of Krylov complexity. This highlights that Krylov complexity serves as a higher-resolution diagnostic tool for models defined on graphs. The distinct dynamical phases characterized by different Krylov complexity scaling could be experimentally probed in atom tweezer setups~\cite{Ebadi2022Quantum}, by measuring out-of-time-order correlations~\cite{hashimoto2017out,swingle2018unscrambling}.

\section{Krylov complexity of free fermions on regular graphs}
To begin, we introduce a free fermion model on a graph, $G(E, V)$. The graph consists of a set of vertices $V$ and a set of edges $E$. Specifically, a regular graph $G_R$ is characterized by each vertex having an identical number of neighbors, indicating that every vertex shares the same degree. Within the framework of the free fermion model, each vertex of a regular graph corresponds to a site, while an edge denotes the tunneling between two such sites. Consequently, the Hamiltonian is represented as follows:
\begin{equation}
    H_{\text{Free}} = \sum_{\{i,j\}\in E} J_{ij}a^{\dag}_ia_j,
    \label{Eq:H_free}
\end{equation}
where $i,j$ labels the vertices, $\{i,j\} \in E$ represent the edges in the graph. 
In this study, we focus on the case with $J_{ij} = J$. 
We consider $N$ sites at half-filling. The tunneling $J_{ij}$ is constrained to regular graphs of degree $d = 2$ and $d = 3$. Although we focus on the fermionic model, we expect the spin model also to exhibit analogous physics through the Jordan-Wigner transformation.

\begin{figure}[htp]
\includegraphics[width=1\linewidth]{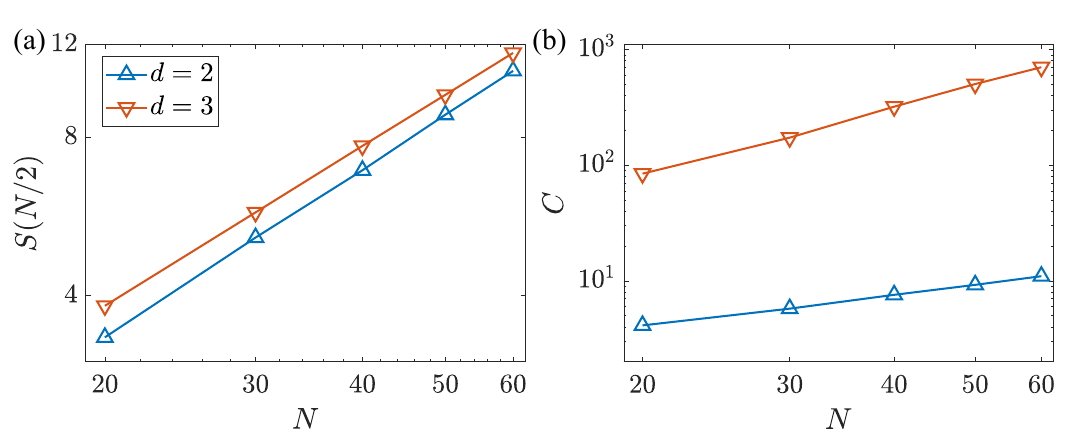}
\caption{The entanglement and Krylov complexity of free fermions on regular graphs. (a), the scaling behavior of entanglement, $S(N/2)$, with respect to the system size $N$. We conduct simulations on $500$ random regular graphs with an evolution time of up to $10^3$ (in units of $\hbar/J$). 
We find entanglement volume law scaling for both $d=2$ and $d = 3$ with $S\sim N$.  (b), the scaling behavior of the Krylov complexity, $C$. For $d = 2$, we average all non-isomorphic graphs. For $d = 3$, we sample $500$ random non-isomorphic graphs. We find $C_{\text{Free}}^{d = 2}\sim N^{0.91(4)}$ for $d =2$ and $C_{\text{Free}}^{d = 3}\sim N^{2.03(3)}$ for $d = 3$.
}
\label{fig:Free}
\end{figure}

In the absence of interactions, fermionic many-body eigenstates are described as Slater-determinant product states. The entanglement entropy of these states is solely influenced by the two-point correlation function denoted as $C_{ij} = \braket{a^{\dag}_ia_j}$~\cite{IngoPeschel2003,Li2016Quantum}. By diagonalizing the correlation matrix within a local subsystem, we obtain eigenvalues $Z_m \in [0,1]$. Therefore, the Von Neumann entropy ($S$) for noninteracting eigenstates can be expressed as $S(l) = -\sum_{m=1}^l (1-Z_m){\rm ln}(1-Z_m) + Z_m{\rm ln}Z_m$. Here, $l$ indicates the number of lattice sites within the subsystem. For our analysis, we set $l = N/2$, with the initial state occupying the first $N/2$ sites with $N/2$ fermions. We track the time evolution of the two-point correlation function, $C_{ij}(t) = \braket{a^{\dag}_i(t)a_j(t)}$, where $a_j(t) = \exp{(-iH_{\text{Free}}t)}a_{j}\exp{(iH_{\text{Free}}t)}$, and evaluate the eigenvalue $Z_m(t)$. The Hamiltonian $H_{\text{Free}}$ is randomly selected from regular graphs with $d = 2$ and $d = 3$. The resulting findings are depicted in Fig.~\ref{fig:Free} (a), illustrating the average $S(N/2)$ over time and various sample graphs. 
By employing log-log plots and linear fitting, we establish scaling laws as $S\sim N^{1.06}$ for $d = 2$ and $S\sim N^{1.00}$ for $d = 3$.
The entanglement for $d = 2$ and $d = 3$ both exhibit volume law scaling.

The free fermionic models defined on regular graphs with $d = 2$ and $d = 3$ belong to the same dynamical phase based on entanglement scaling. Similar conclusions are drawn for interacting systems in Section~\ref{sec:interaction}. However, in some problems, a change in the key parameter from $2$ to $3$ can significantly alter the complexity of the problem. For example, while $2$-SAT is known to belong to the P complexity class, $3$-SAT is NP-complete~\cite{Karp1972}. Another example is that a regular graph state with $d = 2$ can be simulated classically with relative ease, whereas the case with $d = 3$ is much more difficult to simulate classically~\cite{Ghosh2023Complexity}. These examples suggest that complexity may serve as a more informative diagnostic tool than entanglement. Therefore, in the next section, we will examine the complexity of these models.

We proceed to examine the Krylov complexity in regular graphs and consider a many-body Hamiltonian $H$ and an initial local Hermitian operator $O$. The Heisenberg evolution of an operator is described by:
\begin{equation}
    O(t) = e^{iHt}Oe^{-iHt} = \sum_{n = 0}\frac{(it)^n}{n!}\mathcal{L}^nO
\end{equation}
Here, $\mathcal{L}$ represents the Liouville operator, $\mathcal{L}|O) = |[H,O])$. The fictitious operator state is defined as $|O) = \sum_{ij}O_{ij}\ket{i}\bra{j}$, based on an orthonormal basis $\{\ket{i},\ket{j}\}$. The Krylov space is spanned by ${\mathcal{L}^n|O)}$, with its orthonormal basis introduced as ${|O_n)}$ through the orthogonalization of ${\mathcal{L}^n|O)}$. Typically, the Lanczos algorithm is utilized to generate an orthonormal basis of the Krylov space $\{|O_n)\}$. Starting with the initial operator $|O_0) = |O)$, we have $|O_1) = b_1^{-1}\mathcal{L}|O_0)$, where $b_1^2 = (\mathcal{L}O_0|\mathcal{L}O_0)$. For $n\ge2$,
\begin{align}
    |A_n) &= \mathcal{L}|O_{n-1}) - b_{n-1}|O_{n-2}),\\
    b_n^2 &= (A_n|A_n),\\
    |O_n) &= \frac{1}{b_n}|A_n)
\end{align}
The inner product between two operator states is defined as:
\begin{equation}
    (A|B) = \frac{Tr[A^{\dag}B]}{Tr[\mathbf{I}]},
\end{equation}
where $\mathbf{I}$ is the identity matrix.

In the Krylov basis, the Liouvillian operator's matrix representation takes a tridiagonal form, with secondary diagonal elements $b_n$. By expanding the Heisenberg-evolved operator in the Krylov basis:
\begin{equation}
    |O(t)) = \sum_{n = 0}^{D-1} \phi_n(t)|O_n),
\end{equation}
where $D$ is the Krylov dimension. The Heisenberg equation of motion becomes:
\begin{equation}
    -i\partial_t\phi_n(t) = b_n\phi_{n-1}(t) + b_{n+1}\phi_{n+1}(t).
\end{equation}
The initial condition is $\phi_n(0 ) = \delta_{n0}$, where $b_0 = 0$. Therefore, the equation of motion governing $b_n$ can be viewed as a single-particle hopping problem on a semi-infinite chain, with the hopping amplitudes $b_n$. We also have:
\begin{equation}
    \vec{\phi}(t) = e^{-i\mathcal{L}t}\vec{\phi}(0),
\end{equation}
where $\vec{\phi}(t) = (\phi_0(t), \phi_1(t),\cdots)^{T}$. The Krylov complexity is the average position of the propagating packet over the Krylov chain: 
\begin{equation}
    C(t) = \sum_{n = 0}^{D-1} n|\phi(t)|^2.
\end{equation}


We delve into the Krylov complexity for regular graphs with degrees $d = 2$ and $d = 3$. In the subsequent analysis, we solely consider non-isomorphic regular graphs, as isomorphic ones exhibit identical Krylov complexities (Supplementary). In our study, we investigate the operator growth~\cite{bhattacharyya2023operator,Parker2019Universal,caputa2024krylov} of particle number operators, $O_i = a^{\dag}_ia_i$, where $i$ ranges from 1 to $N$. The corresponding results are illustrated in Fig.~\ref{fig:Free} (b). We averaged over different initial operators and regular graphs, as well as the time-averaged Krylov complexity, which occurs only when the Krylov complexity reaches a plateau. More details of the Krylov dynamics are provided in the Supplementary. Our analysis involves studying the scaling laws associated with the Krylov complexity. Through log-log plots and linear fitting, we have established scaling relationships for the Krylov complexity. Specifically, for the Krylov complexity of free fermions on regular graphs of $d = 2$ ($C_{\text{Free}}^{d=2}$), the scaling law is $C_{\text{Free}}^{d=2} \sim N^{0.91(4)}$. Similarly, for the Krylov dimension  of free fermions on regular graphs of $d = 3$ 
($C_{\text{Free}}^{d=3}$), 
the scaling law is $C_{\text{Free}}^{d=3} \sim N^{2.03(3)}$. The $(3)$ and $(4)$ indicate errors caused by numerical calculations, such as the gradual loss of orthogonality during the orthogonalization process.

The entanglement of regular graphs with $d=2$ and regular graphs with $d=3$ both adhere to the volume law, implying that they fall within the same phase. The entanglement properties can not distinguish between them. However, their Krylov complexity scaling laws diverge significantly, positioning them in different complexity phases. This observation suggests that the dynamical phases existing on regular graphs undergo enrichment through the scaling of Krylov complexity. Despite showcasing identical volume-law entanglement characteristics, dynamical phases may reveal distinct Krylov complexity patterns, as illustrated by our developed free fermion model on regular graphs.

\section{Krylov complexity of interacting fermions on regular graphs}
\label{sec:interaction}
We also investigate the interacting fermions model.
The Hamiltonian is defined as follows~\cite{Amin2024Solvable}:
\begin{equation}
    H = \sum_{\{ij\}\in E}^{N} J_{ij} (a^{\dag}_i a_j + n_i n_j),
\end{equation}
where $n_i = a^{\dag}_i a_i$ represents the particle number operator. Here, the tunneling and interaction are defined using the same graphs, i.e., sharing the same factor $J_{ij}$. As a result, the interacting Hamiltonian is still determined by a regular graph and the interaction does not break the structure of regular graphs. 

\begin{figure}[htp]
\includegraphics[width=1\linewidth]{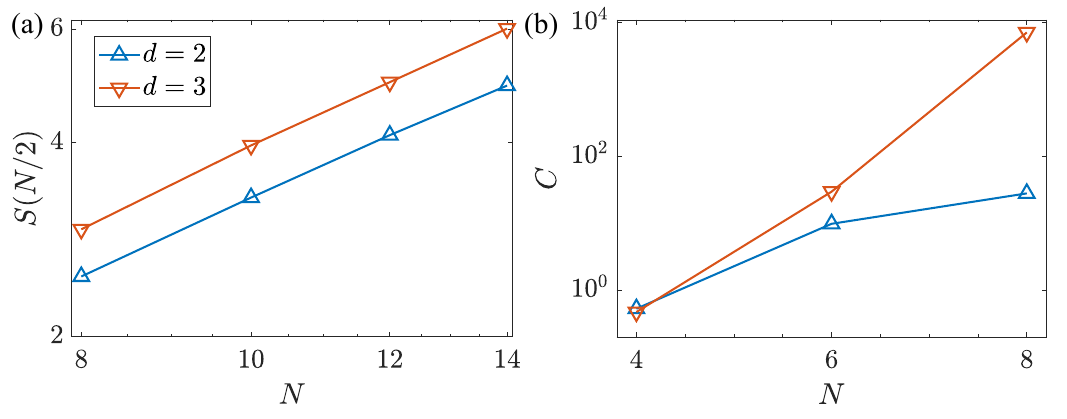}
\caption{ The entanglement and Krylov complexity for interacting fermions.
 (a), the scaling behavior of entanglement, $S (N/2)$, to the system size $N$.  Simulations are conducted on $500$ random regular graphs with an evolution time of up to $10^3$ (in units of $\hbar/J$). 
 We find that the entanglement of regular graphs with $d = 2$ and $d = 3$ scales with $S(N/2)\sim N$, adhering to the volume law. (b), the Krylov complexity for systems of small size. Here, we calculate the average across all non-isomorphic graphs.}
\label{fig:ManyBody}
\end{figure}

The initial states are also half-filling. We track the time evolution of the initial state and evaluate the Von Neumann entropy $S(N/2)$. The Hamiltonian $H$ is randomly selected from regular graphs with $d = 2$ and $d = 3$. The resulting findings are depicted in Fig~\ref{fig:ManyBody} (a), illustrating the average $S(N/2)$ over time across various sample graphs. We observe that both the entanglement of regular graphs with $d = 2$ and $d = 3$ scales with $S(N/2)\sim N$, satisfying the volume laws.

For interacting fermions, the computation cost for calculating the Krylov complexity grows exponentially with the system size. Our numerical simulations are restricted to relatively small systems, as depicted in Fig~\ref{fig:ManyBody} (b). We average over all non-isomorphic graphs and initial operators. Evidently, the Krylov complexity of regular graphs with $d = 2$ and $d = 3$ exhibits exponential growth. Notably, the regular graph with $d = 3$ demonstrates a quicker rate of increase. Based on findings from analyses involving smaller sizes, the Krylov complexity scaling for regular graphs with $d = 3$ could be $\sim \exp{(N)}$, whereas for $d = 2$, it could follow an exponential growth pattern of $\sim \exp{(\sqrt{N})}$.

Whether examining a free fermion model or an interacting fermion model, their entanglement adheres to the volume law, denoted as $S \sim N$. Nevertheless, in terms of Krylov complexity, the complexity of these models differs significantly. Even with the same category of fermion model, if placed on distinct regular graphs, their complexities will exhibit marked divergences. Consequently, the dynamical phases experience enrichment through Krylov complexity.

\section{Krylov dimension of free fermions on regular graphs}
We have established the scaling laws of Krylov complexity in free fermions and examined the Krylov complexity for interacting fermions of small sizes. In this section, our focus lies on the Krylov dimension. In some cases, the Krylov dimension can also reflect the complexity of the system, such as whether it is chaotic or non-chaotic~\cite{Parker2019Universal,barbon2019evolution,caputa2024krylov}. Through the application of reasonable assumptions, we are able to efficiently compute the Krylov dimension. Consequently, viewing the scenario from the perspective of the Krylov dimension permits a quantifiable comprehension of the scaling laws governing free fermions and enables an estimation of the scaling laws associated with interacting fermions.

The distinctive behaviors in the Krylov complexity of regular graphs with $d=2$ and $d=3$ (Eq~(\ref{Eq:H_free})) can be understood through their connectivity. As depicted in Fig.~\ref{fig:SketchMap}, regular graphs with $d = 2$ consist of disconnected loops, while those with $d = 3$ are almost fully interconnected. This structural contrast significantly impacts the Krylov complexity. Both the Hamiltonian $H_{\text{free}}$ and the local operator $O_i$ are quadratic operators, and their commutators are also quadratic operators. Consequently, the Krylov dimension scales approximately as $N^2$. This analysis is consistent with Krylov complexity scaling laws for $d=3$. The free fermion Hamiltonian of regular graphs with $d = 2$ 
($H_{\text{Free}}^{d = 2}$) can be represented as $H_{\text{Free}}^{d = 2} = \sum_l H_l$, where $l$ represents distinct disconnected loops, each with a length of $L_l$, and $\sum_l L_l = N$. The operators $H_l$ mutually commute, i.e., $[H_l, H_{l'}]=0$. If our initial operator $O_i$ is positioned on a single loop $l$, the Krylov dimension is determined by the loop's length $L_l$, not the system size $N$, due to the commuting property of $H_l$. Various initial operators $O_i$ may localize to different loops, each with its distinct length. This implies that the Krylov dimension of $d=2$ is much smaller than that of $d=3$. To evaluate the Krylov dimension of a regular graph with $d = 2$, we need to average the selection of initial operators based on the probability $L_l/N$.

More quantitatively, we analyze the Krylov dimension of free fermions on regular graphs by averaging over their disconnected subgraphs. In our theoretical framework, we make a crucial assumption: the Krylov dimension of a free Hamiltonian defined on a graph is proportional to the square of the loop length where the initial operator is located. In a regular graph of d = 3, there is at least one path connecting any two arbitrary vertices, and we consider its loop length to be N. This assumption is deemed reasonable, given that our focus is on extracting scaling laws rather than precisely determining the Krylov dimension. Furthermore, our findings indicate that the Krylov dimension for a free Hamiltonian defined on a loop scale proportionally to the square of the loop lengths (Supplementary).

The theory consists of two primary components: one involves counting the number of non-isomorphic regular graphs with $d = 2$, while the other entails evaluating the Krylov dimension for each non-isomorphic graph. Counting the number of non-isomorphic graphs is akin to solving the problem of partitioning $N$ into parts $\geq 3$, represented as $N = \sum_{l = 1}^M L_l,~3\leq L_l$, where $1 \leq M \leq \lfloor \frac{N}{3} \rfloor$ denotes the number of integers. For instance, $M=2$ corresponds to partitioning $N$ vertices into two subsets. The independent ways of partition, denoted by $(L_1, L_2)$, are $[(3, N-3), (4, N-4), \cdots, (\lfloor N/2\rfloor, N - \lfloor N/2\rfloor)]$, with a total number of $C(N,M=2,I=3) = \lfloor \frac{N}{2}\rfloor - 2$. Here, $I = 3$ is the minimal integer in the decompositions, and the corresponding contribution to the Krylov dimension is denoted as $D_{\text{Free}}(N, M=2, I=3)$,
\begin{equation}
    D_{\text{Free}}(N,M=2,I=3) = \sum_{3\leq k\leq \lfloor \frac{N}{2}\rfloor } k^2\frac{k}{N} + (N-k)^2\frac{N-k}{N},
\end{equation}
where \(\frac{k}{N}\) denotes the probability that the initial local operator is in the loop with a loop length of $k$. The factor $k^2$ and $(N-k)^2$ represent the corresponding the Krylov dimension. For any $M = m$, we derive a recurrence relationship as follows (Appendix):
\begin{align}
    C(N, m, I=3) &= \sum_{3\leq k\leq \lfloor \frac{N-k}{m-1} \rfloor} C(N-k, m-1, k), \\
    D_{\text{Free}}(N, m, I=3) &= \sum_{3\leq k\leq \lfloor \frac{N-k}{m-1}\rfloor} k^2\frac{k}{N} + D(N,m-1,k). 
    \label{Eq:FreeRecurrence}
\end{align}
Utilizing this recurrence, we can theoretically estimate the Krylov dimension of regular graphs with \(d = 2\), denoted as \(D^{d =2}_{\text{Theory}}\), which is given by:
\begin{equation}
    D^{d =2}_{\text{Theory}} = \frac{\sum_M  D_{\text{Free}}(N, M, I =3)}{\sum_M C(N, M, I = 3)}.
\end{equation}

\begin{figure}[htp]
\includegraphics[width=1\linewidth]{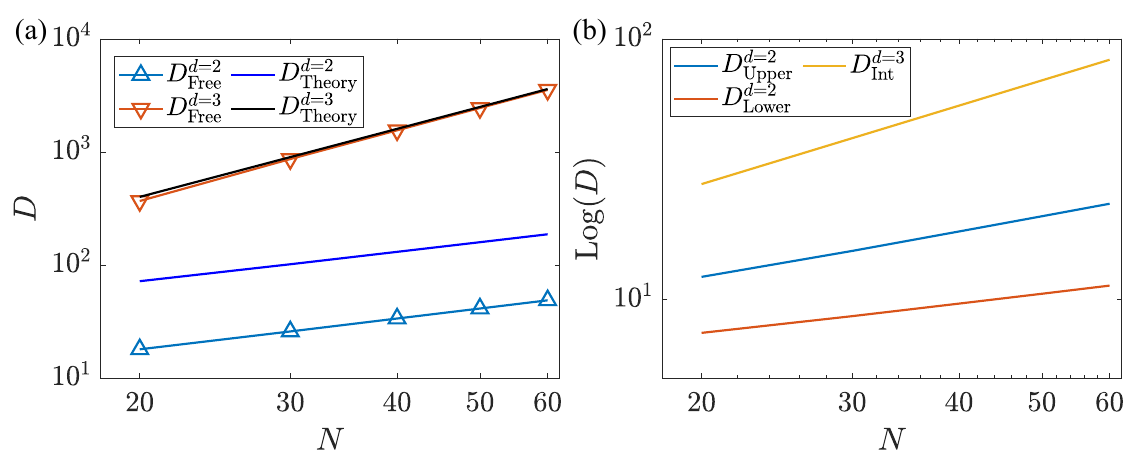}
\caption{The  Krylov dimension for free fermions and interacting fermions. (a), the theoretical and numerical results of the Krylov dimension for free fermions. Regarding regular graphs with $d = 2$, we calculated averages across all non-isomorphic graphs and 500 randomly selected non-isomorphic graphs for $d = 3$. Through linear fitting, we derived the following results: $D^{d = 2} \sim N^{0.91(4)}$, $D^{d = 2}_{\text{Theory}} \sim N^{0.90}$, $D^{d = 3}_{\text{Theory}} \sim N^{2.00}$, and $D^{d = 3} \sim N^{2.03(3)}$. (b), the theoretical results of the Krylov dimension for interacting fermions. In the case of regular graphs with $d = 3$, the Krylov dimension $D^{d=3}_{\text{Int}}$ scales approximately as $4^N$. Concerning regular graphs with $d = 2$, we esitmate the upper limit as $D^{d=2}_{\text{Upper}} \sim 4^{N^{0.59}}$ and the lower limit as $D^{d=2}_{\text{Lower}} \sim 4^{N^{0.38}}$.}
\label{fig:KrylovDimenison}
\end{figure}

The results presented in Fig.~\ref{fig:KrylovDimenison} (a) show that the theoretical scaling law for $d = 2$ approximates $D^{d = 2}_{\text{Theory}} \sim N^{0.90}$. Comparatively, our numerical calculations yield $D^{d = 2} \sim N^{0.91(4)}$. Although the theoretical calculation for the regular graph with $d = 2$ suggests a slightly higher Krylov dimension than the numerical result, their scaling exponents are in strong agreement. As we emphasized in our initial assumptions, our focus is primarily on the scaling relationship of the Krylov dimension rather than its precise value. For regular graphs with $d = 3$, we find $D^{d = 3}_{\text{Theory}} \sim N^{2.00}$, while our numerical result is $D^{d = 3} \sim N^{2.03(3)}$. Here, our theoretical and numerical results align perfectly, both in terms of the scaling relationship and the magnitude of the Krylov dimension, further validating the soundness of our hypotheses. 

Upon analysis, we observed that in the free fermion model, both Krylov complexity and Krylov dimension exhibit similar scaling behavior. Qualitatively, Krylov complexity can be interpreted as the average position of the propagating packet along the Krylov chain, where the length of the chain corresponds to the Krylov dimension. Over long-time evolution, the initial packet will spread across the entire Krylov chain in our model, making it reasonable to expect that the long-time average of Krylov complexity shares the same scaling as the Krylov dimension. In summary, within the system under study, the scaling of the Krylov dimension is consistent with the scaling of Krylov complexity.

\section{Krylov dimension of interacting fermions on regular graphs}
Owing to interactions, numerical computations restrict the Krylov dimension to smaller sizes. Even with these constraints, it is evident that the Krylov dimension scales exponentially as the system size grows (Supplementary). Moreover, our theory effectively establishes the correct scaling correlation for the Krylov dimension and Krylov complexity concerning free fermions. Consequently, we delve into the analysis of the Krylov dimension for interacting fermions by expanding our analytical framework from the non-interacting scenario, instead of relying solely on small-scale numerical computations. Drawing inspiration from free fermions, we establish a scaling relationship for the Krylov dimension to function as an estimation for the Krylov complexity. A practical assumption we make asserts that the Krylov dimension of an interacting Hamiltonian, defined on a graph scale, scales as $4^L$, where $L$ represents the loop length where the scrutinized operator is initially positioned.

Based on this assumption, the Krylov dimension of interacting fermions on regular graphs with $d = 3$ ($D^{d =3}_{\text{Int}}$) follows to the scaling law $D^{d =3}_{\text{Int}}\sim 4^N$, as almost all regular graphs with $d = 3$ are connected. In the case of interacting fermions on regular graphs with $d = 2$, we employ the same approach used in the free scenario. This involves the enumeration of non-isomorphic regular graphs and the subsequent evaluation of the Krylov dimension for each unique graph. In our interacting Hamiltonian setup, both tunneling and interaction are determined by the same graphs, implying that the count of non-isomorphic graphs remains unchanged between interaction and free cases, still denoted as $C(N, M, I)$. Regarding the Krylov dimension for each non-isomorphic graph in the interaction scenarios, we establish a recursive relationship. The corresponding Krylov dimension $D_{\text{int}}(N, m, I=3)$ is outlined as follows:
\begin{equation}
   D_{\text{Int}}(N, m, I=3) = \sum_{3\leq k\leq \lfloor \frac{N-k}{m-1}\rfloor} 4^k\frac{k}{N} + D(N,m-1,k),
\end{equation}
where $4^k$ is the Krylov dimension that the initial local operator is in the loop with a loop length of $k$. The average Krylov dimension across non-isomorphic graphs computes as $D^{d =2}_{\text{Int}} = \sum_M D_{\text{Int}}(N, M, I =3)/\sum_M C(N, M, I = 3)$. By calculating $D^{d = 2}_{\text{Int}}$, we estimate the Krylov dimension, yielding a scaling law of $D^{d = 2}_{\text{Upper}}\sim4^{N^{0.59}}$, as depicted in Fig~\ref{fig:KrylovDimenison} (b). This scaling law serves as an upper bound for the Krylov dimension, as $4^{L}$ is the maximum dimension of the operator space for an interacting fermion Hamiltonian defined on a loop of length $L$.

Here, we establish a lower bound by computing the average loop length $L^{d = 2}$. We formulate a recursive relationship for the loop length $L$ (Appendix):
\begin{equation}
    L(N, m, I=3) = \sum_{3\leq k\leq \lfloor \frac{N-k}{m-1}\rfloor} k\frac{k}{N} + L(N,m-1,k),
\end{equation}
where $k$ represents the loop length housing the initial local operator. Through averaging across non-isomorphic graphs, the average loop length is determined as $L^{d = 2} = \sum_M L(N, M, I=3)/\sum_M C(N, M, I=3)$. Based on our assumption, the Krylov dimension is $B\times 4^{L_i}$, where $B$ is a constant number and $L_i$ signifies the respective loop length. We infer that according to the arithmetic means surpassing or equating the geometric mean, $B\times 4^{L^{d = 2}}$ will be less than or equal to the average of $B\times 4^{L_i}$ across various non-isomorphic graphs. Consequently, by computing $4^{L^{d = 2}}$, we arrive at a lower bound, leading to $D^{d = 2}_{\text{Lower}} \sim 4^{L^{d = 2}}$. Upon evaluating $L^{d = 2}$, we estimate the Krylov dimension, resulting in $D^{d = 2}_{\text{Int}}\sim 4^{0.38}$, as illustrated in Fig~\ref{fig:KrylovDimenison}(b).

Given that nearly all non-isomorphic regular graphs with $d = 3$ are connected graphs, the average loop length equates to the system size $N$. Consequently, the lower limit of the Krylov dimension for regular graphs with $d = 3$ stands at $4^N$. Simultaneously, its upper boundary aligns with $4^N$, leading to a scaling law of $D^{d = 3}_{\text{Int}}\sim 4^N$, depicted in Fig~\ref{fig:KrylovDimenison} (b). These semianalysis scaling laws will be extrapolated to small-scale systems and compared with the numerical results we obtained at small scales. We found that these estimates are still consistent with the numerical calculations, detailed discussions in Supplementary.

\section{Out-of-time-order correlation}
To probe the distinctive dynamical phases on the graph enriched by the complexity, we propose to examine the quantum dynamics of atom tweezer arrays, where non-local couplings can be engineered by moving the atoms individually~\cite{barredo2016atom,endres2016atom}. 
It is worth noting that direct measurement of the Krylov complexity is practically infeasible. 
Nonetheless, the dynamical phases of regular graphs with $d=2$ and $d=3$ having distinctive complexity also produce qualitatively different OTOC behaviors, which can then be measured in experiments~\cite{Li2017Measuring,Xiao2021Information,Green2022Experimental,landsman2019verified}. We thus expect the complexity enriched dynamical phases to be accessible to current experiments. Additionally, the OTOC is the lower bound of Krylov complexity for any chaotic system, specifically, $\lambda_L\leq 2\beta$. The $\lambda_L$ is the Lyapunov exponent and $\beta$ is the growth rate of Lanczos coefficients ~\cite{Parker2019Universal}. In the SYK model, equality can be achieved. Thus, the measurement of OTOC also can indirectly detect the Krylov complexity.

The OTOC is a time-dependent function defined by an averaged double-commutator as
\begin{equation}
    C(t) = \braket{[O(t),Q]^{\dag}[O(t),Q]}.
\end{equation}
If $O$ and $Q$ are both hermitian and unitary, 
\begin{equation}
    C(t) = 2(1 - Re(F(t))),
\end{equation}
where $F(t) = \braket{O(t)^{\dag}Q^{\dag}O(t)Q}$. The OTOC can be detected in experiments and used to distinguish between regular graphs with degrees 2 and 3. The most significant disparity between degrees 2 and 3 is evident in the structure of the regular graphs. A degree-2 regular graph consists of disconnected loops, while a degree-3 regular graph is nearly connected. This contrast can be discerned using the OTOC. In a degree-2 regular graph, the operators $O(t)$ and $Q$ commute if their support belongs to different loops, resulting in $C(t) = 0$. However, this scenario is rare in the case of degree 3.

\begin{figure}[htp]
\includegraphics[width=1\linewidth]{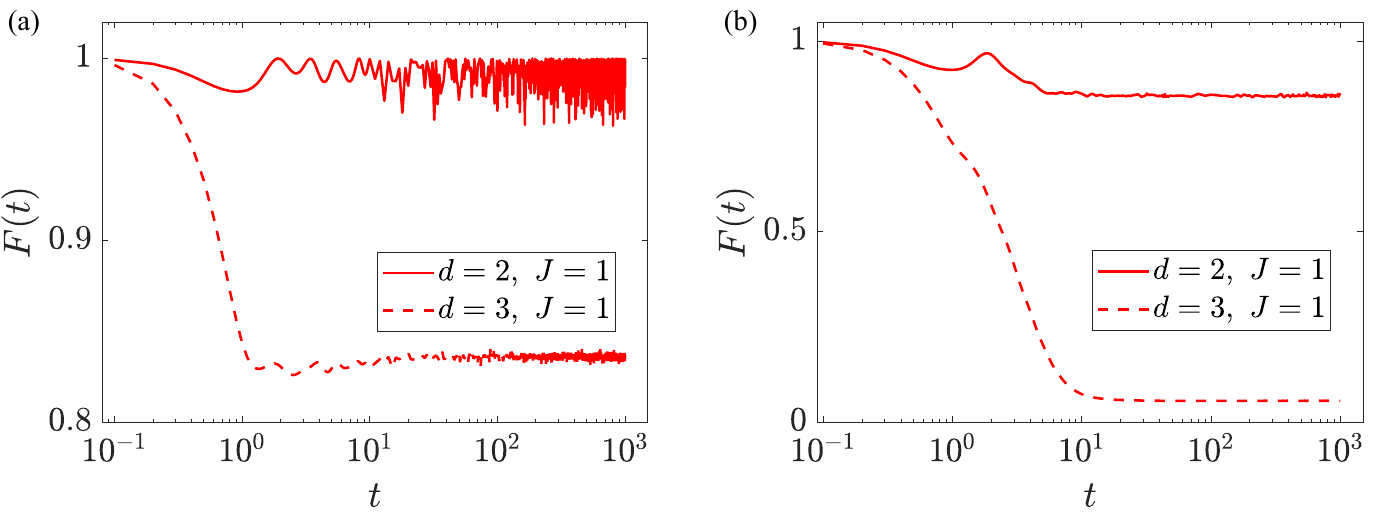}
\caption{The numerical results of OTOC. (a), the plot of $F(t)$ as a function of $t$ for free fermions. (b), the plot of $F(t)$ as a function of $t$ for interacting fermions. The initial operators considered are $O = 2a_1^{\dag}a_1 - \mathbf{I}$ and $Q = 2a_{N/2}^{\dag}a_{N/2}- \mathbf{I}$. 
In the free case, the system size is $N = 20$. We sample all non-isomorphic graphs for regular graphs of $d = 2$ and 500 non-isomorphic graphs for regular graphs of $d = 3$. In the interacting case, the system size is $N = 14$, and we sample all non-isomorphic regular graphs for both $d = 2$ and $d = 3$.}
\label{fig:OTOC}
\end{figure}

Here, we consider the operators $O = 2a_1^{\dag}a_1 - \mathbf{I}$ and $Q = 2a_{N/2}^{\dag}a_{N/2}- \mathbf{I}$ to ensure that $Q$ and $O$ are unitary and Hermitian. We investigate the quantity $F(t)$ averaged over non-isomorphic graphs, as depicted in Fig~\ref{fig:OTOC}. Fig~\ref{fig:OTOC} (a) shows the results for free fermions with $N = 20$. 
As we initially anticipated, for $ d = 2 $, the structure of the corresponding graph in the system is disconnected, resulting in most of the different non-isomorphic graphs contributing an OTOC of 0, leading to $ F(t) $, being close to 1. In contrast, for $ d = 3 $, nearly all non-isomorphic graphs are connected, which contributes to a non-zero OTOC, causing $ F(t) $ to decrease over time. However, $ F(t) $ stabilizes at a relatively large value, approximately 0.82, indicating that the system lacks scrambling ability. Since we are considering free fermions, where the growth rate of the Lanczos coefficient $\beta $ is 0~\cite{Parker2019Universal}, this implies that the system's Lyapunov exponent is also 0.

Figure~\ref{fig:OTOC} (b) presents the results for interacting fermions with $ N = 14 $. For regular graphs with $ d = 2 $, $ F(t) $ remains close to 1, similar to the behavior observed in the free case. However, for $ d = 3 $, $ F(t) $ decreases and stabilizes around 0, indicating that the system undergoes scrambling. By applying exponential fitting, we can extract the Lyapunov exponent $ \lambda_L = 0.31(2) $. Consequently, $ 0.31\leq 2\beta $, and the Krylov complexity increases exponentially at least as $ \sim \exp{(0.31t)} $~\cite{Parker2019Universal}. As Krylov complexity increases exponentially over time, the propagating packet on the Krylov chain rapidly spreads throughout the entire chain. In this case, the saturation value of Krylov complexity is determined by the size of the Krylov dimension, which explains why the Krylov dimension and Krylov complexity exhibit similar scaling behavior. Whether in the free or interacting scenario, the OTOC results for $ d = 2 $ and $ d = 3 $ exhibit significant differences. This notable contrast in experimental observations underscores the distinct dynamical phases in regular graphs with $ d = 2 $ and $ d = 3 $.

\section{Conclusion and Outlook}
We have investigated both entanglement and Krylov complexity for fermions on regular graphs with degrees $d = 2$ and $d = 3$. The non-local tunneling and interactions in these graph models may be experimentally realized using $^6$Li atoms confined by optical tweezers~\cite{Two2022Yan,qiu2020precise}. Entanglement scales as $S\sim N$ on both types of regular graphs, regardless of whether $d = 2$ or $d = 3$. In contrast, the Krylov complexity exhibits distinct behaviors on these graphs. In the absence of interactions, the Krylov dimension scales as $D^{d=2}_{\text{Free}}\sim N^{0.91(4)}$ for $d = 2$, and as $D^{d=3}_{\text{Free}}\sim N^{2.03(3)}$ for $d = 3$. When interactions are present, the Krylov dimension scales as $D^{d=2}_{\text{Int}}\sim 4^{N^{\alpha}}$ for $d = 2$, where $0.38 \leq \alpha \leq 0.59$, and as $D^{d=3}_{\text{Int}}\sim 4^N$ for $d = 3$.

To interpret the complex behavior of Krylov complexity, we have developed an analytic theory that accurately reproduces the intricate scaling of the Krylov dimension, showing excellent quantitative agreement with numerical simulations. Based on entanglement scaling, the models under consideration belong to the same dynamical phase. However, Krylov complexity reveals distinct scaling behavior in these models, suggesting that Krylov complexity serves as a higher-resolution diagnostic tool for models on regular graphs. It is capable of identifying more dynamical phases than entanglement alone. The differences in fermion dynamics on graphs with $d = 2$ and $d = 3$, and their varying associated complexities, can be probed through out-of-time-order correlators (OTOCs), which are accessible in current experimental setups. The potential phase transitions between dynamical phases of different complexities warrant further investigation.

{\it Acknowledgement.---}
This work is supported by National Key Research and Development Program of China (2021YFA1400900), National Natural Science Foundation of China (11934002), and Shanghai Municipal Science and Technology Major Project (Grant No. 2019SHZDZX01).

\bibliography{references}

\clearpage

\begin{widetext} 
\renewcommand{\theequation}{S\arabic{equation}}
\renewcommand{\thesubsection}{S-\arabic{subsection}}
\renewcommand{\thefigure}{S\arabic{figure}}
\renewcommand{\thetable}{S\arabic{table}}
\setcounter{equation}{0}
\setcounter{figure}{0}
\setcounter{table}{0}

\newpage

\begin{center} 
{\Huge \bf Supplementary Materials} \\
\end{center} 
\section{The Krylov complexity}
\subsection{The definition of Krylov complexity}
We consider a Hamiltonian $H$ and an initial local Hermitian operator $O$. For any operator $O = \sum_{ij}O_{ij}\ket{i}\bra{j}$, defined on an orthonormal basis $\ket{i},\ket{j}$, the corresponding operator state is denoted as $|O) = \sum_{ij}O_{ij}\ket{i}\bra{j}$, and the inner product between two operator states is defined as:
\begin{equation}
    (A|B) = \frac{Tr[A^{\dag}B]}{Tr[\mathbf{I}]},
\end{equation}
where $\mathbf{I}$ is the identity matrix. The Heisenberg evolution of the operator $O$ is given by:
\begin{equation}
    O(t) = e^{iHt}Oe^{-iHt} = \sum_{n = 0}\frac{(it)^n}{n!}\mathcal{L}^nO,
\end{equation}
where $\mathcal{L}|O) = |[H,O])$. Equivalently, the Heisenberg evolution can be expressed as:
\begin{equation}
     |O(t)) = \sum_{n = 0}\frac{(it)^n}{n!}\mathcal{L}^n|O).
\end{equation}
The Krylov space is spanned by ${\mathcal{L}^n|O\rangle}$. Typically, the Lanczos algorithm is utilized to generate an orthonormal basis of the Krylov space $\{|O_n)\}$. Starting with the initial operator $|O_0) = |O)$, we have $|O_1) = b_1^{-1}\mathcal{L}|O_0)$, where $b_1^2 = (\mathcal{L}O_0|\mathcal{L}O_0)$. For $n\ge2$,
\begin{eqnarray}
    |A_n) &=& \mathcal{L}|O_{n-1}) - b_{n-1}|O_{n-2}),\\
    b_n^2 &=& (A_n|A_n),\\
    |O_n) &=& \frac{1}{b_n}|A_n)
\end{eqnarray}
It is known that the original Lanczos algorithm features a significant numerical instability: the construction of each Krylov element involves the two previous ones, causing errors due to finite-precision arithmetic to accumulate rapidly, leading to a loss of orthogonality in the Krylov basis during numerical computations. Residual overlaps between Krylov elements grow exponentially with the iteration number $n$, rendering the Lanczos coefficients unreliable after a few iterations. To address this issue, we adopt the following strategies~\cite{barbon2019evolution}:
\begin{enumerate}
    \item $|O_0) = \frac{1}{\sqrt{(O|O)}}|O)$.
    \item For $n\ge 1$: compute $|A_n) = \mathcal{L}|O_{n-1})$.
    \item Re-orthogonalize $|A_n)$ explicitly with respect to all previous Krylov elements: $|A_n)\rightarrow|A_n) - \sum_{m = 0}^{n-1}|O_{m})(O_m|A_n)$.
    \item Repeat step 3.
    \item Set $b_n = \sqrt{(A_n|A_n)}$.
    \item if $b_n = 0$ = stop; otherwise set $|O_n) = \frac{1}{b_n}|A_n)$ and go to step 2.
\end{enumerate}

$|O_n)$ involves an n-nested commutator with the Hamiltonian, and the operator becomes more complex and nonlocal with increasing orders of $n$. Therefore, the order $n$ can serve as a measure of operator complexity, motivating the definition of Krylov complexity:
\begin{equation}
    C(t) = \sum_{n = 0}^{D-1}n|(O(t)|O_n)|^2.
\end{equation}
Here, $D$ represents the dimension of the Krylov space. In the Krylov basis, the Liouvillian operator's matrix representation takes a tridiagonal form, with secondary diagonal elements $b_n$. By expanding the Heisenberg-evolved operator in the Krylov basis:
\begin{equation}
    |O(t)) = \sum_{n = 0}^{K-1} \phi_n(t)|O_n),
\end{equation}
the Heisenberg equation of motion becomes:
\begin{equation}
    -i\partial_t\phi_n(t) = b_n\phi_{n-1}(t) + b_{n+1}\phi_{n+1}(t).
\end{equation}
The initial condition is $\phi_n(0 ) = \delta_{n0}$, where $b_0 = 0$. Therefore, the equation of motion governing $b_n$ can be viewed as a single-particle hopping problem on a semi-infinite chain, with the hopping amplitudes $b_n$. We also have:
\begin{equation}
    \vec{\phi}(t) = e^{-i\mathcal{L}t}\vec{\phi}(0),
\end{equation}
where $\vec{\phi}(t) = (\phi_0(t), \phi_1(t),\cdots)^{T}$. The Krylov complexity is the average position of the propagating packet over the Krylov chain: 
\begin{equation}
    C_D(t) = \sum_{n = 0}^{D-1} n|\phi(t)|^2.
\end{equation}
\subsection{The numerical results of Krylov dimension}
We also explore the dimension of Krylov space in many-body scenarios. However, the exponentially large Krylov space, approximately $4^N$, renders numerical calculations impractical. Here, we present some results for small sizes, as depicted in Fig~\ref{fig:SupManyBodyKrylovDimension}. We average overall non-isomorphic graphs and initial operators. It's evident that both the Krylov space dimension $D$ increase exponentially. We extrapolate the fitted Krylov dimension to small sizes, and the corresponding results are shown in Fig~\ref{fig:SupManyBodyKrylovDimension} (a) and (b). In the case of a regular graph with $d = 3$, $D^{d=3}_{\text{Theory}}$ serves as an upper bound for $D^{d=3}$, and particularly at $N=8$, $D^{d=3}_{\text{Theory}}$ aligns closely with $D^{d=3}$. Our theoretical and numerical findings for free fermions with $d = 3$ are entirely consistent. This suggests that as the system size grows, we expect $D^{d=3}_{\text{Theory}}$ and $D^{d=3}$ will continue to closely align. On a regular graph with $d = 2$, $D^{d=2}_{\text{Upper}}$ surpasses $D^{d=2}$, maintaining its role as an upper limit. Notably, $D^{d=2}_{\text{Upper}}$ and $D^{d=2}$ are almost parallel. It's worth recalling that for free fermions with $d = 2$, our theory accurately predicts the scaling relationships. Therefore, we anticipate that as system sizes increase, the scaling behaviors of $D^{d=2}_{\text{Upper}}$ will mirror those of $D^{d=2}$. Despite the current observation that our estimated lower bound, $D^{d=2}_{\text{Lower}}$, is larger than $D^{d=2}$, based on their growth patterns, we foresee that with larger sizes, $D^{d=2}_{\text{Lower}}$ will indeed fall below $D^{d=2}$.

\begin{figure}[htp]
\includegraphics[width=1\linewidth]{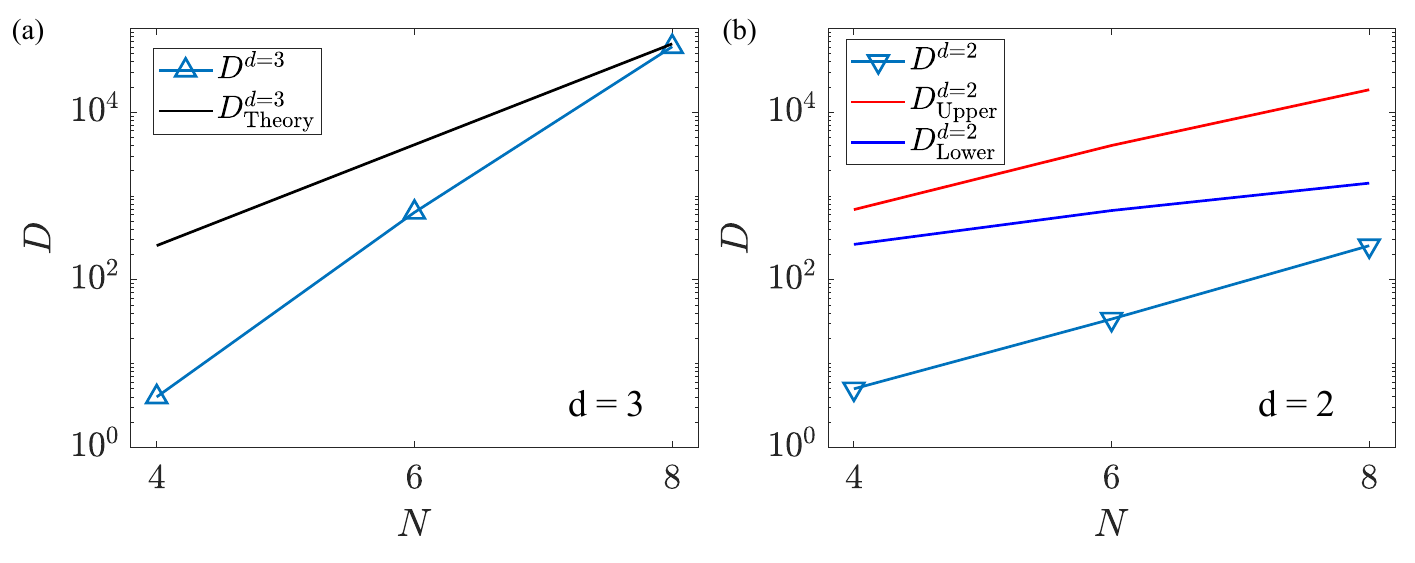}
\caption{The numerical results of interacting fermions on graphs. (a) the plot of Krylov dimension $D$ as a function of system size $N$ for $d = 3$. (b) the plot of Krylov dimension $D$ with respect to system size $N$ for $d = 2$. For both regular graphs with $d = 2$ and $d = 3$, we average over all non-isomorphic graphs and initial operators.}
\label{fig:SupManyBodyKrylovDimension}
\end{figure}

\section{The theoretical framework}
We examine the free fermion Hamiltonian, $H_{\text{free}} = \sum_{ij}J_{ij}a^{\dagger}_ia_j$, and the local particle density operators $O_i = a^{\dagger}_ia_i$. Both $H_{\text{free}}$ and $O_i$ are quadratic operators, and their commutators are also quadratic operators. Thus, the Krylov space is spanned by quadratic operators, and the Krylov dimension scales as the square of the system size $N$, $D \sim N^2$. We also consider that for free fermions on regular graphs with $d = 2$, the Hamiltonian can be rewritten as $H^{d = 2}_{\text{free}} = \sum_l H_l$, where $l$ labels distinct disconnected loops, each with length $L_l$, and $\sum_l L_l = N$. The Hamiltonian $H_l$ defined on the disconnected subgraphs commutes with each other. The commutator $ [\sum_l H_l, O_i] = [H_{l_i}, O_i]$, where $l_i$ represents the loop in which the vertex $i$ is located. Thus, the Krylov dimension is determined by the loop length $L_{l_i}$ rather than the system size $N$. Additionally, we focus on the scaling laws rather than the exact values of the dimension of the Krylov space. Based on the above observations, we make a reasonable assumption that the dimension of the Krylov space of a free Hamiltonian defined on a graph is proportional to the square of the loop length where the initial operator is located. We also validate this assumption numerically. We create loops with varying lengths and examine their Krylov space dimension, as depicted in Fig~\ref{fig:SupLoopAssmuption}. Through logarithmic plotting and linear fitting, we find that the dimension $D$ scales as $D\sim N^{2.00}$, consistent with our assumption.
\begin{figure}[htp]
\includegraphics[width=0.5\linewidth]{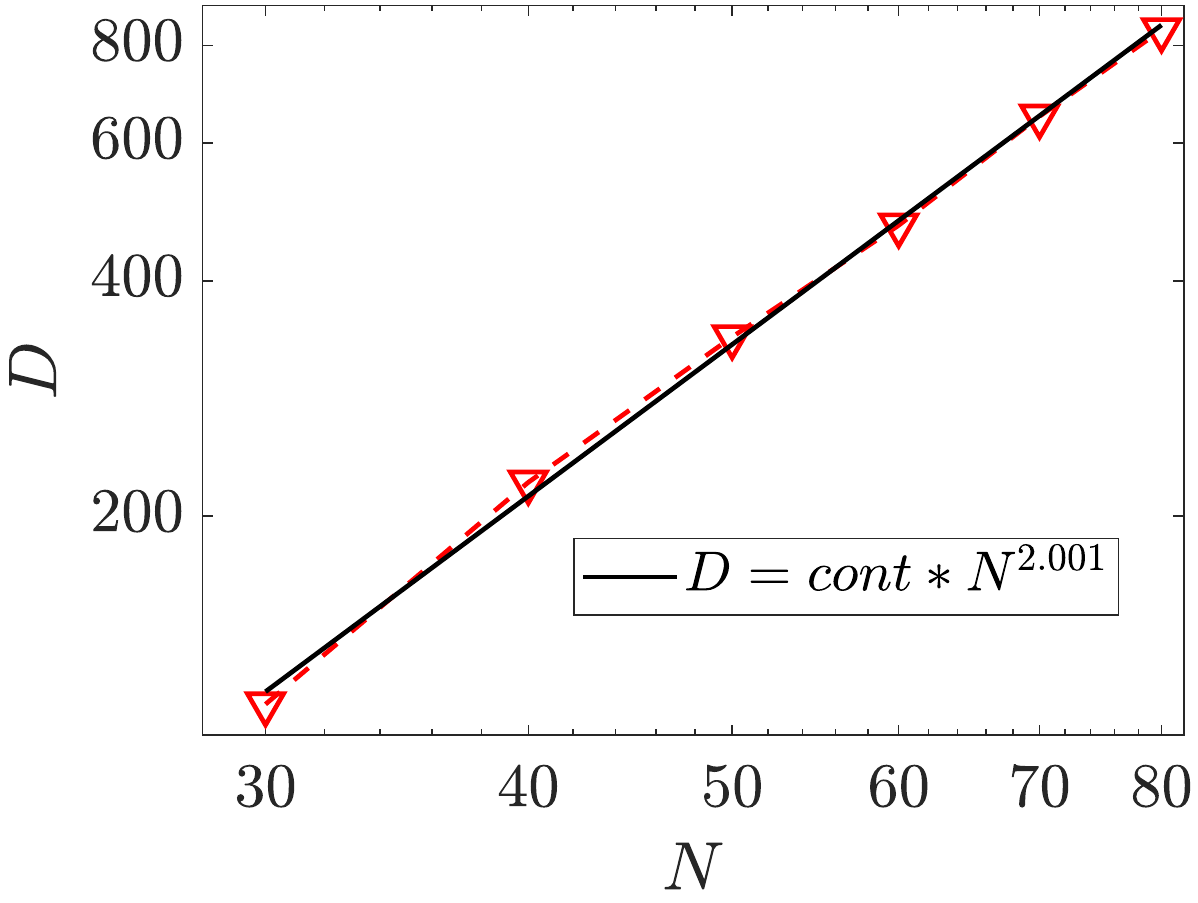}
\caption{The dimension of Krylov space for loops with varying lengths.}
\label{fig:SupLoopAssmuption}
\end{figure}

Based on this assumption, considering that the non-isomorphic graphs of $d = 3$ are almost connected, the scaling laws for regular graphs of $d = 3$ are $D_{\text{Free}}^{d = 3} \sim N^2$. Next, we derive the scaling laws for regular graphs of $d = 2$. To achieve this, we need to count the number of non-isomorphic graphs, denoted as $C$, and determine Krylov dimension for each graph, denoted as $D$. The problem of determining the number of non-isomorphic graphs is equivalent to the problem of partitioning $N$ into parts $\geq 3$, such that $N = \sum_{l = 1}^M L_l$, where $L_l \geq 3$, and $1 \leq M \leq \lfloor \frac{N}{3} \rfloor$ represents the number of integers or the number of loops in regular graphs.

For $M = 1$, there is only one decomposition, resulting in the number of decompositions $C(N, M=1, I=3) = 1$, and the corresponding contribution to the Krylov space dimension is $D_{\text{Free}}(N, M=1, I=3) = N^2$.
\begin{eqnarray}
    C(N, M=1, I=3) = 1, \\
    D_{\text{Free}}(N, M=1, I=3) = N^2
\end{eqnarray}
Here, $I$ is the minimum integer in the decompositions. For $M = 2$, where $N = n_1 + n_2$, there are $\lfloor \frac{N}{2}\rfloor - 2$ decompositions, represented by $[(3, N-3), (4, N-4), \cdots, (\lfloor N/2\rfloor, N - \lfloor N/2\rfloor)]$. The corresponding contribution to the Krylov dimension is denoted as:
\begin{eqnarray}
 C(N,M=2,I=3) &=& \lfloor \frac{N}{2}\rfloor - 2, \\
 D_{\text{Free}}(N,M=2,I=3) &=& \sum_{3\leq k\leq \lfloor \frac{N}{2}\rfloor } k^2\frac{k}{N} + (N-k)^2\frac{N-k}{N}.
\end{eqnarray}
Here, $k$ represents the length of a loop, and $k^2$ is the corresponding dimension of the Krylov space. $\frac{k}{N}$ is the probability that the initial operator $O_i$ is located in this loop.

For $M = 3$, our strategy involves dividing $N$ into two integers, $L_1$ and $L_2$, just as we do in $M = 2$, and then further dividing $n_2$ into 2 integers. For example, $N = 15$ can be divided into $L_1 = 3, L_2=12$ or $L_1 = 4, L_2= 11$, then $L_2 = 3 + 9$ or $L_2 = 4 + 7$. However, there is repetitive counting, such as $L_2 = 4 + 8$ and $L_2 = 3 + 8$. The key point is to avoid this issue. We take $N = 15$ as an example to elaborate on our method. The following table shows all the decompositions of $N = 15$.
\begin{center}
\begin{tabular}{ c c c }
 3 & 3 & 9 \\
 3 & 4 & 8 \\  
 3 & 5 & 7 \\ 
 3 & 6 & 6 \\   
 4 & 4 & 7 \\   
 4 & 5 & 6 \\   
 5 & 5 & 5   
\end{tabular}
\label{Tab:decomposition}
\end{center}
First, we set $L_1 = 3$, and $N-L_1$ is divided into 2 integers $[(3,9), (4,8), (5,7), (6,6)]$. Next, for $L_1 = 4$, to avoid repetition, the minimal integers of the decompositions $I$ should satisfy $I=4$. Similarly, for $L_1 = 5$, $I=5$. Thus, 
\begin{eqnarray}
 C(N,M=3,I=3) &=& \sum_{3\leq k \leq \lfloor \frac{N-k}{2}\rfloor}C(N-k, M=2, k) \\
 D_{\text{Free}}(N,M=3,I=3) &=& \sum_{3\leq k\leq \lfloor \frac{N-k}{2}\rfloor} k^2\frac{k}{N} + D_{\text{Free}}(N,M=2,k).
\end{eqnarray}
Similarly, we derive the recurrence for any $M$:
\begin{eqnarray}
 C(N,M,I=3) &=& \sum_{3\leq k \leq \lfloor \frac{N-k}{M-1}\rfloor}C(N-k, M-1, k) \\
 D_{\text{Free}}(N,M,I=3) &=& \sum_{3\leq k\leq \lfloor \frac{N-k}{M-1}\rfloor} k^2\frac{k}{N} + D_{\text{Free}}(N,M-1,k).
\end{eqnarray}
Based on the recurrence relation, the dimension of Krylov space $D_{\text{Free}}^{d = 2} = \sum_M D_{\text{Free}}(N, M, I =3)/\sum_M C(N, M, I =3)$.

The theory can also be extended to include interacting fermions on graphs. In this scenario, the Krylov space is not solely spanned by quadratic operators but also encompasses other many-body operators. Consequently, we have revised the assumption made in the free case, now positing that the Krylov dimension of an interacting Hamiltonian defined on a graph is proportional to $4^L$, where $L$ is the loop length where the initial operator is located. Similarly, we can derive a recurrence relation for the interacting case:
\begin{equation}
     D_{\text{Int}}(N,M,I=3) = \sum_{3\leq k\leq \lfloor \frac{N-k}{M-1}\rfloor} 4^k\frac{k}{N} + D_{\text{Int}}((N,M-1,k).
\end{equation}
The Krylov dimension of interacting fermions $D_{\text{Int}}^{d = 2} = \sum_M D_{\text{Int}}(N, M, I =3)/\sum_M C(N, M, I =3)$.

This method can also be utilized to calculate the loop length. In the case of regular graphs with $d = 3$, where the graphs are almost entirely connected, the loop length equals $N$. However, for regular graphs with $d = 2$, the loop length of a non-isomorphic graph is determined by:
\begin{equation}
     L(N,M,I=3) = \sum_{3\leq k\leq \lfloor \frac{N-k}{M-1}\rfloor} k\frac{k}{N} + L((N,M-1,k).
\end{equation}
$K$ is the length of a loop. The average loop length of interacting fermions in the case of $d = 2$ is $L^{d = 2} = \sum_M L(N, M, I =3)/\sum_M C(N, M, I =3)$ and the scaling laws of Krylov dimension is $\sim 4^{L^{d =2}}$. Based on the recurrence relation, we can generate all possible decompositions. For example, the integer $9$ can be decomposed into $[(9),(3,6),(4,5),(3,3,3)]$. The total decompositions $\sum_M C(N, M, I =3) = 4$ and the corresponding Krylov dimension for free case $D_{\text{Free}}(N, M, I =3) = 9^2\times\frac{9}{9} + 3^2\times\frac{3}{9} + 6^2\times\frac{6}{9} + 4^2\times\frac{4}{9} + 5^2\times\frac{5}{9}+  + 3^2\times\frac{3}{9} + 3^2\times\frac{3}{9}+ 3^2\times\frac{3}{9}$. $D_{\text{Free}}^{d = 2} = \sum_M D_{\text{Free}}(N, M, I =3)/\sum_M C(N, M, I =3) = 31.5$.

\section{Krylov dynamics with longer time}
Within the main text, we delve into the time-averaged Krylov complexity. Here, we present the Krylov complexity dynamics illustrated in Fig~\ref{fig:SupKCLongTime} (a) and (b). Our base operators consist of particle number operators, $O_i = a^{\dag}_ia_i$, covering indices from 1 to $N$. We average these initial operators and non-isomorphic graphs, effectively averaging the Hamiltonian. The Krylov dynamics exhibit three distinct phases: linear growth, exponential growth, and a plateau phase. When contrasting the outcomes for $d = 3$, the Krylov complexity in regular graphs with $d = 2$ shows smaller values and oscillations. This phenomenon arises from the makeup of regular graphs with $d = 2$ that comprise loops of varying lengths, where Hamiltonians corresponding to different loops commute. The Krylov dimension primarily hinges on the loop length rather than the system's overall size, leading to notable fluctuations in Krylov complexity. The constrained Krylov dimension in turn restricts the growth of Krylov complexity. Conversely, in regular graphs with $d = 3$, nearly all non-isomorphic graphs are connected.

Furthermore, we explored the scaling relationships over an extended duration, delineated in Fig~\ref{fig:SupKCLongTime} (c). We specifically considered the time-averaged complexity only upon reaching a plateau. By employing a log-log plot, we can unveil the scaling relationship. For $d = 2$, we observe $C\sim N^{0.90(5)}$, while for $d = 3$, the trend follows $C\sim N^{2.01(2)}$. These findings are consistent with the main text within the margin of error.
\begin{figure}[htp]
\includegraphics[width=1\linewidth]{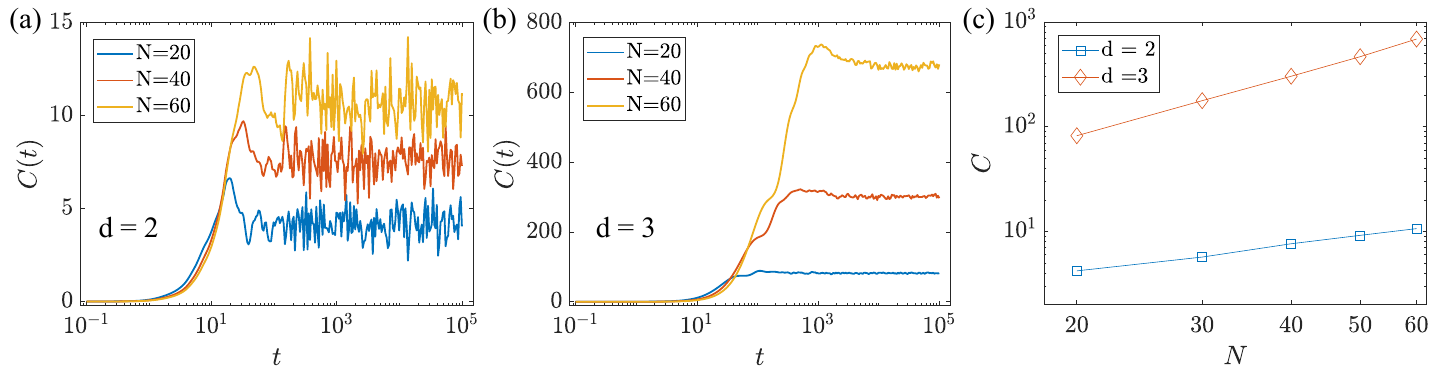}
\caption{Krylov dynamics with longer time. (a), the Krylov dynamics for regular graphs of $d = 2$. We average the initial operators and all non-isomorphic graphs. (b), the Krylov dynamics for regular graphs of $d = 3$. We average the initial operators and 500 non-isomorphic graphs. (c), the scaling laws of Krlov complexity. The time average occurs when the complexity reaches a plateau. The corresponding scaling ara $C\sim N^{0.9(5)}$ for $d = 2$ and $C\sim N^{2.01(2)}$ for $d = 3$.}
\label{fig:SupKCLongTime}
\end{figure}

\section{The discussion about the non-isomorphic graphs}
In the main text, we focus exclusively on non-isomorphic graphs. Let's explore this concept further. We begin with the free Hamiltonian $H_{\text{free}} = \sum_{\{i,j\}\in E} J_{ij}a^{\dag}_ia_j$, defined over the operator basis ${a^{\dag}_i, a_j}$. Here, $J$ denotes our Hamiltonian. In the case of isomorphic regular graphs, typically, there exists a permutation difference between them. Consequently, the original commutator $[H,O_i]$ is transformed into $[PHP^{\dag},O_i]$, where $P$ represents a permutation transformation. Since $[PHP^{\dag}, O_i] = P[H, P^{\dag}O_iP]P^{\dag}$ and $P^{\dag}O_iP \neq O_i$, the dimension of the Krylov space for isomorphic graphs may vary. However, it's important to note that the initial operator $O_i$ is diagonal, and $O_j = P^{\dag}O_iP$ is also diagonal. Therefore, when summing over all initial operators, $\sum_i O_i = \sum_i P^{\dag}O_iP$, since $\sum_i O_i = \mathbf{I}$. This implies that the dimension of the Krylov space for isomorphic graphs is the same based on the average of the initial operators. 

For interacting fermions on regular graphs, represented by the Hamiltonian $H = \sum_{\{ij\}\in E}^{N}J_{ij}(a^{\dag}_ia_j + n_in_j)$, isomorphic graphs are related by $J\rightarrow J^{\prime} = PJP^{\dag}$. Here, $J$ and $J^{\prime}$ correspond to two Hamiltonians $H$ and $H^{\prime}$ in the many-body basis, respectively. We observe that the two Hamiltonians differ only by relabelling the indices ${i,j}\rightarrow {i^{\prime}, j^{\prime}}$. Hence, similarly, the dimension of the Krylov space for isomorphic graphs remains the same, based on the average of the initial operators.

To validate the aforementioned arguments, we conducted a numerical investigation of the dimension of the Krylov space for isomorphic graphs. We selected a regular graph $G_R$ and generated an isomorphic graph $G^{\prime}_R$ by applying a random permutation to the adjacency matrix. The dimension of the Krylov space for both $G_R$ and $G^{\prime}_R$, denoted as $D(G_R)$ and $D(G^{\prime}_R)$ respectively, was computed. We then averaged the results over all initial operators $O_i,~i\in [1, N]$. The corresponding outcomes are depicted in Fig~\ref{fig:SupVerifyIsomorhic}. In our analysis, we sampled 40 non-isomorphic regular graphs, with the x-coordinate $G$ representing different non-isomorphic graphs. We defined the ratio $R = D(G_R)/D(G^{\prime}_R)$ to quantify the difference between the two isomorphic graphs. From the numerical results, it was evident that the ratio $R$ equaled exactly 1. This implies that isomorphic graphs exhibit the same dimension of Krylov space, validating our theoretical analysis. Similar conclusions were reached for interacting fermions. However, it's worth noting that due to computational constraints, we were only able to investigate very small system sizes in many-body cases, and the number of non-isomorphic graphs was limited. For instance, When $ N =12, $the number of non-isomorphic regular graphs of $d = 2$ was only 2. Therefore, we did not present the numerical results for interacting fermions.

\begin{figure}[htp]
\includegraphics[width=0.5\linewidth]{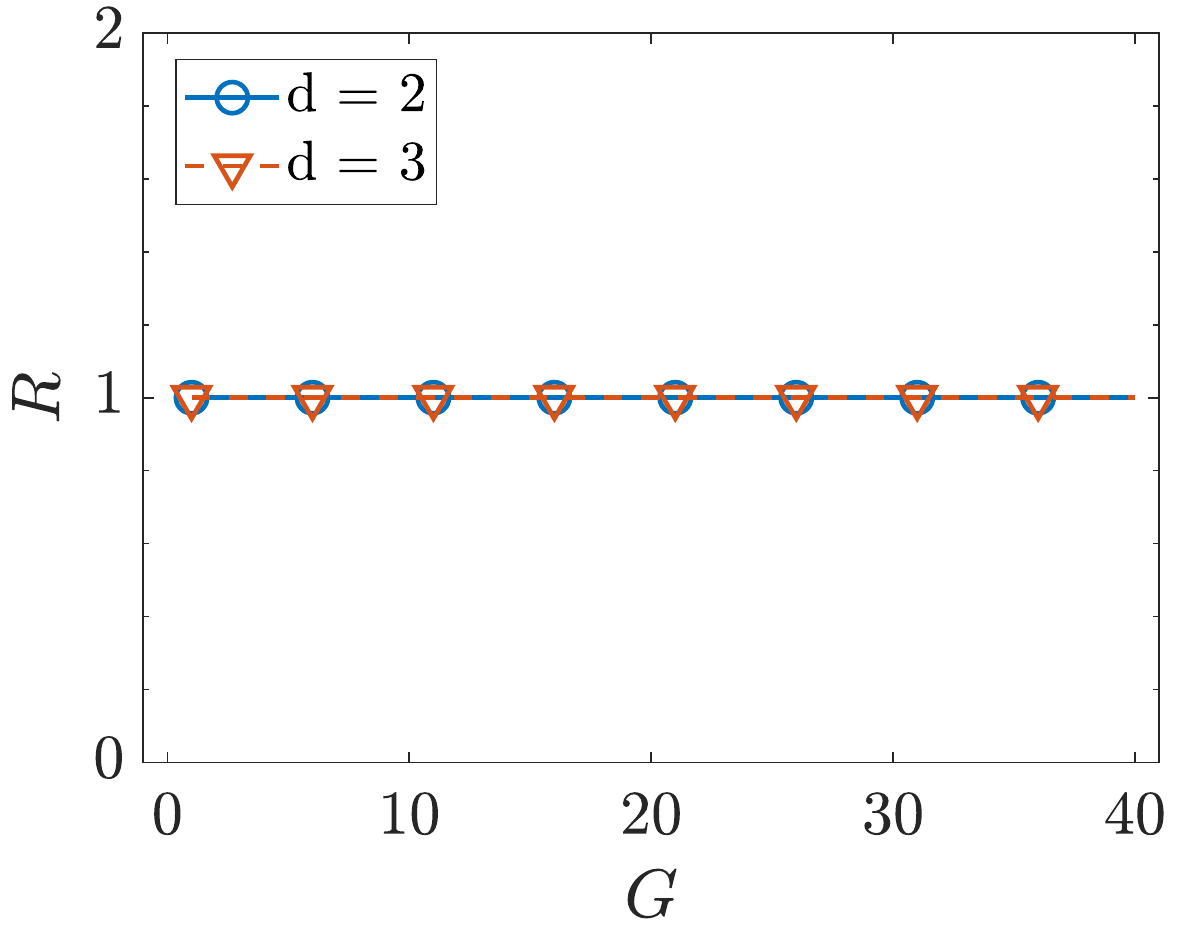}
\caption{The dimension of the Krylov space for isomorphic regular graphs. The x-coordinate $G$ represents different non-isomorphic regular graphs. We sampled 40 non-isomorphic regular graphs, as this encompasses the total number of non-isomorphic regular graphs of $d = 2$, which is approximately 40. The results are averaged over all initial operators.}
\label{fig:SupVerifyIsomorhic}
\end{figure}
\section{The effect of disorder}
We also explore the effects of disorder on free fermions in graphs. The Hamiltonian is defined as follows:
\begin{equation}
    H_{\text{dis}} = \sum_{\{i,j\}\in E} J_{ij}a^{\dag}_ia_j + \sum_{i\in G_R} W_ia^{\dag}_ia_i.
\end{equation}
Here, we introduce on-site disorder to the Hamiltonian, with $W_i$ representing the disorder strength, uniformly distributed in the range $[-W, W]$. We examine the Krylov complexity, as illustrated in Fig ~\ref{fig:SupDisorder}. In this analysis, we sample 200 disorder instances with $N = 50$. An intriguing observation emerges: for regular graphs of $d = 3$, the Krylov complexity decreases with increasing disorder strength. This outcome is expected, as disorder typically induces localization. However, for $d = 2$, the Krylov complexity initially increases and then decreases with disorder. We speculate that this behavior arises from the interplay between symmetry and disorder.
\begin{figure}[htp]
\includegraphics[width=0.5\linewidth]{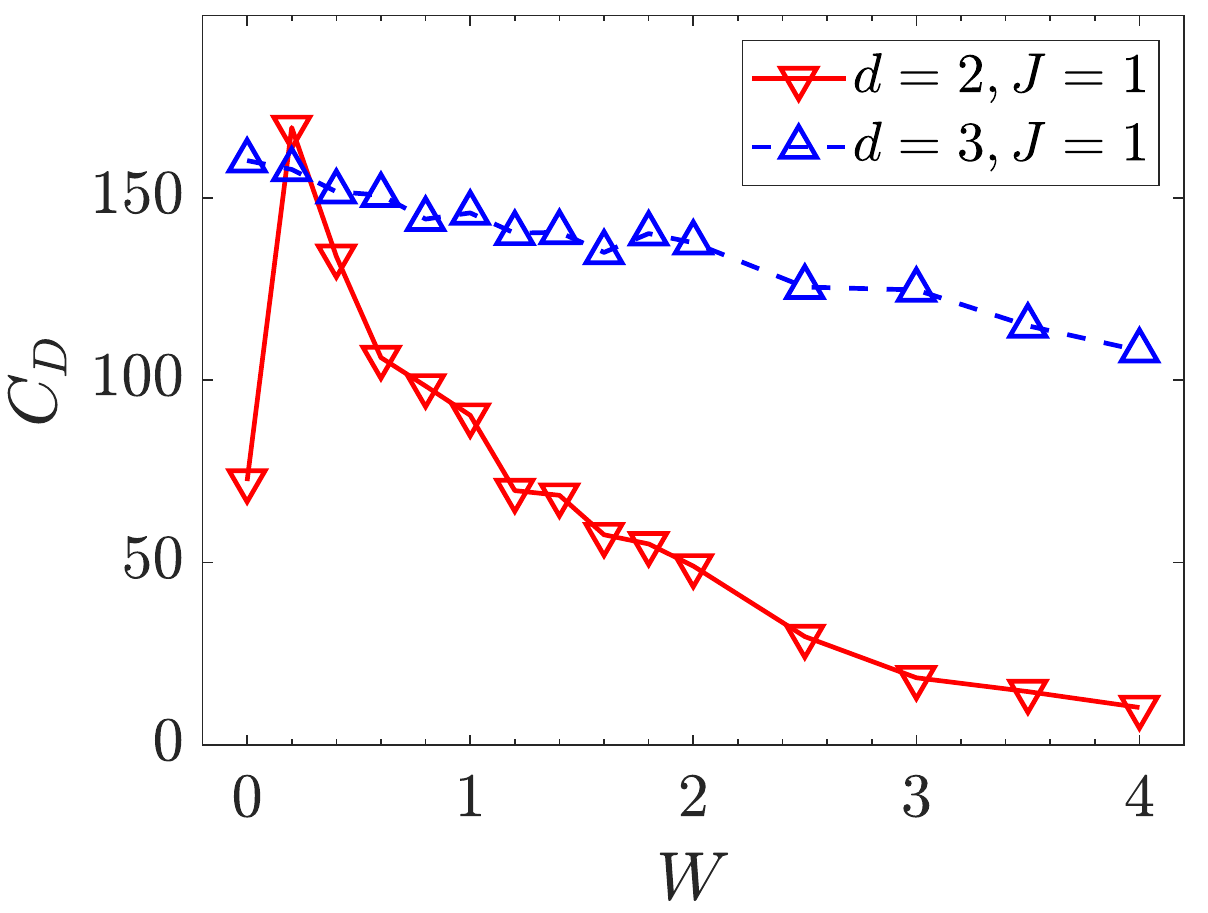}
\caption{The Krylov complexity of disorder-free fermions on graphs. We consider a system size of $N = 50$ and sample 200 disorder configurations.}
\label{fig:SupDisorder}
\end{figure}



\end{widetext}
\end{document}